%% file: AA.tex
\documentclass[letterpaper,11pt]{article}
\usepackage{times}
\usepackage{amsfonts}

\usepackage{epic,eepic,amsmath,latexsym,amssymb,color}
\usepackage{ifthen,graphics,epsfig,subfigure,fullpage}
\usepackage[english]{babel}
\usepackage{times}
\usepackage{url}
\usepackage{array}
\usepackage{color}
\usepackage{colortbl}

\usepackage{pgf}
\usepackage{tikz}
\usetikzlibrary{arrows,matrix, calc, fit}

\usepackage{rotating}

\usepackage{pgfplots}

\usepackage{comment}

\usepackage{amsmath}
\usepackage{amsmath}
\usetikzlibrary{patterns}

\begin{document}

  \def\startCirc#1{\tikz[remember picture,overlay]\path node[inner sep=0, anchor=south] (st) {#1} coordinate (start) at (st.center);}%
  \def\endCirc#1{\tikz[remember picture,overlay]\path node[inner sep=0, anchor=south] (en) {#1} coordinate (end) at (en.center);%
    \begin{tikzpicture}[overlay, remember picture]%
      \path (start);%
      \pgfgetlastxy{\startx}{\starty}%
      \path (end);%
      \pgfgetlastxy{\endx}{\endy}%
      \pgfmathsetlengthmacro{\xdiff}{\endx-\startx}%
      \pgfmathsetlengthmacro{\ydiff}{\endy-\starty}%
      \pgfmathtruncatemacro{\xdifft}{\xdiff}%
      \pgfmathsetmacro{\xdiffFixed}{ifthenelse(equal(\xdifft,0),1,\xdiff)}%
      \pgfmathsetmacro{\angle}{ifthenelse(equal(\xdiffFixed,1),90,atan(\ydiff/\xdiffFixed))}%
      \pgfmathsetlengthmacro{\xydiff}{sqrt(abs(\xdiff^2) + abs(\ydiff^2))}%
      \path node[draw,rectangle,red, rounded corners=2mm, rotate=\angle, minimum width=\xydiff+4ex, minimum height=2.5ex] at ($(start)!.5!(end)$) {};%
    \end{tikzpicture}%
  }

\newlength {\squarewidth}
\renewenvironment {square}
{
\setlength {\squarewidth} {\linewidth}
\addtolength {\squarewidth} {-12pt}
\renewcommand{\baselinestretch}{0.75} \footnotesize
\begin {center}
\begin {tabular} {|c|} \hline
\begin {minipage} {\squarewidth}
\medskip
}{
\end {minipage}
\\ \hline
\end{tabular}
\end{center}
}



\newtheorem{definition}{Definition}
\newtheorem{theorem}{Theorem}
\newtheorem{lemma}{Lemma}
\newtheorem{corollary}{Corollary}
\newtheorem{property}{Property}
\newtheorem{observation}{Observation}
\newcommand{\note}[1]{\noindent\textcolor{red}{{\fontfamily{phv}\selectfont NOTE: #1}}}
\newcommand{\toto}{xxx}
\newenvironment{proofT}{\noindent{\bf
Proof }} {\hspace*{\fill}$\Box_{Theorem~\ref{\toto}}$\par\vspace{3mm}}
\newenvironment{proofL}{\noindent{\bf
Proof }} {\hspace*{\fill}$\Box_{Lemma~\ref{\toto}}$\par\vspace{3mm}}
\newenvironment{proofC}{\noindent{\bf
Proof }} {\hspace*{\fill}$\Box_{Corollary~\ref{\toto}}$\par\vspace{3mm}}

\newcounter{linecounter}
\newcommand{\linenumbering}{\ifthenelse{\value{linecounter}<10}
{(0\arabic{linecounter})}{(\arabic{linecounter})}}
\renewcommand{\line}[1]{\refstepcounter{linecounter}\label{#1}\linenumbering}
\newcommand{\resetline}[1]{\setcounter{linecounter}{0}#1}
\renewcommand{\thelinecounter}{\ifnum \value{linecounter} >
9\else 0\fi \arabic{linecounter}}

\renewcommand{\note}[1]{\noindent\textcolor{red}{{\fontfamily{phv}\selectfont NOTE: #1}}}
\newcommand{\SB}[1]{\noindent\textcolor{red}{{\fontfamily{phv}\selectfont SB-NOTE: #1}}}
\newcommand{\AD}[1]{\noindent\textcolor{blue}{{\fontfamily{phv}\selectfont AD-NOTE: #1}}}
\newcommand{\MP}[1]{\noindent\textcolor{green}{{\fontfamily{phv}\selectfont MP-NOTE: #1}}}
\newcommand{\ST}[1]{\noindent\textcolor{olive}{{\fontfamily{phv}\selectfont ST-NOTE: #1}}}
\newcommand{\vir}[1]{``#1''}
\newcommand{\cured}[1]{\emph{cured}}
\newcommand{\correct}[1]{{\em correct}}

%
\title{Approximate Agreement under Mobile Byzantine Faults}

\author{Silvia Bonomi$^\star$, Antonella Del Pozzo$^\star$$^\dagger$, Maria Potop-Butucaru$^\dagger$, S\'ebastien Tixeuil$^\dagger$\\~\\
$^\star$Sapienza Universit\`{a} di Roma,Via Ariosto 25, 00185 Roma, Italy\\
\texttt{\{bonomi, delpozzo\}}$@$dis.uniroma1.it\\
$^\dagger$Sorbonne Universit\'{e}s, UPMC, LIP6-CNRS 7606 -- 4, Place Jussieu, Paris, France\\
\texttt{\{maria.potop-butucaru, sebastien.tixeuil\}}$@$lip6.fr}


\date{ }
\maketitle

\begin{abstract}

In this paper we address Approximate Agreement problem in the Mobile Byzantine faults model. Our contribution is three-fold.
First, we propose the \emph{the first mapping} from the  existing variants of Mobile Byzantine models to the Mixed-Mode faults model.This mapping further help us to prove the correctness of class MSR (Mean-Subsequence-Reduce) Approximate Agreement algorithms   in the Mobile Byzantine fault model, and is of independent interest. Secondly,  we prove \emph{lower bounds} for solving Approximate Agreement under all existing Mobile Byzantine faults models. Interestingly,  these lower bounds are different from the static bounds. Finally, we propose matching upper bounds.

Our paper is the \emph{first to link the Mobile Byzantine Faults models and the Mixed-Mode Faults  models}, and we advocate that a similar approach can be adopted in order to prove the correctness of other classical distributed building blocks (\emph{e.g.} agreement, clock synchronization, interactive consistency etc)  under Mobile Byzantine Faults model.  
\end{abstract}


\section{Introduction}

\input{introduction}

\section{Related Works}

\input{relatedWork}

\input{contributions}
\section{System Model and Problem Definition}
\label{section:systemmodel}
\input{systemModel}



\section{Mapping Mobile Byzantine Faults to Mixed-model Faults Model}
\label{sec:background}

\input{background}
\section{MSR under Mobile Byzantine Faults}

\input{mappingProofs}

\section{Lower Bounds}

\input{lowerBounds}

\section{Conclusions}
\input{conclusions}

\bibliographystyle{IEEEtran}
\bibliography{latex8,references1}

\end{document}

%% file: introduction.tex
The emergent area of sensor networks or mobile robot networks revived recently the research on one of the most studied building blocks of distributed computing: \emph{Approximate Agreement}~ \cite{BPT09c,BPT10j,Charron-BostFN15,LHW14,SV15,TV13,TV14a,TV14b,VTL12}. Indeed, gathering environmental data such as temperature or atmospheric pressure, or synchronizing clocks in large scale sensor networks, typically do not require perfect agreement between participating nodes. Also, requiring autonomous mobile robots to gather at some specific location \emph{e.g.} to communicate or to setup a new task  tolerates a difference in the final robot positions after gathering. This is due to the robots physical size. Accepting a predetermined difference in the agreement process permits to avoid many impossibility results occurring in the perfect agreement case. 

The Approximate Agreement problem \cite{DLPS86,KA94,Lynch96} is nevertheless complex to solve in systems prone to Byzantine faults. 
In sensor networks, sensors may not transmit their values or may transmit erroneous values due to permanent or temporary failures. In mobile autonomous robot networks, some robots may move in the opposite direction as the one intended due to hardware malfunction of buggy software. In both cases the signals (transmitted data, or perceived position) sent by the faulty participants may have a tremendous impact on the approximated value that is computed by the correct ones. The main criterium for evaluating the complexity of Approximate Agreement in a particular setting is by providing the maximum proportion of participants that may exhibit arbitrary behavior in any system execution (w.r.t. the total number of participants). The other participants are considered to \emph{never} deviate from their specification.

The problem becomes even more difficult to solve when faults are mobile. That is, when the faulty behavior may impact different participants over time. For example, in sensor or mobile robot networks, the possibility of intermittent external perturbations (\emph{e.g.} magnetic fields) may affect different processes of the network at various moments during system execution. Participants that are located in such affected areas may exhibit Byzantine behavior. Obviously, in these systems the definition of a "correct" and "corrupted" process is not trivial since a correct process may be corrupted temporarily afterwards, while a corrupted process may behave again according to its specifications, once the external perturbation ceased. When faults are mobile, every process may exhibit Byzantine behavior in a given system execution. So, complexity criteria that were valid for the static case must be redesigned from scratch in systems with dynamically evolving faults.

\paragraph{Our Contribution.}
  This paper considers the Approximate Agreement problem, where processes start with real values from some interval, and are required to converge, after a sequence of voting rounds, to a set of values that are within $\epsilon$ of each other, where $\epsilon$ denotes a (strictly) positive real number. When the environment is prone to Byzantine faults, faulty processes may exhibit arbitrary behavior and in particular may play against the correct ones in order to prevent their convergence.
We address the Approximate Agreement problem under the Mobile Byzantine Faults model, where an adversary controls Byzantine agents and moves them from one process to another. When such an agent is located at a process, this process may behave arbitrarily (and even maliciously). We consider a round-based synchronous computational model where the movements of the agents are synchronized with the change of rounds. 
This paper studies conditions to achieve Approximate Agreement in the four existing synchronous Mobile Byzantine Faults models, that differ in the diagnosis capabilities of processes, \emph{e.g.}, when processes can diagnose their failure state (that is, they are aware that the mobile agent has left them), and when processes cannot self-diagnose.
We  prove {\emph lower bounds} (that are different from the static case) on the number of correct processes, $n$, that is necessary to achieve Approximate Agreement in the presence of $f$ Mobile Byzantine Faults (that is, $f$ agents) for each of the four models. Then we extend the correctness proof of the MSR (Mean-Subsequence-Reduce) class of Approximate Agreement algorithms,  \cite{KA94}, to the Mobile Byzantine faults model.
Our correctness proof makes use for the fist time in this context of a {\emph mapping} between the  Mobile Byzantine Faults models and the Mixed-Mode Faults model  \cite{KA94} composed of asymmetric (classical Byzantine), symmetric and benign static faults. The benign faults are self-incriminating (immediately self-evident to all non faulty processes). The behavior of symmetric faults is perceived identically to all correct processes, while the asymmetric faults have a totally arbitrary behavior. 
Our mapping  is of independent interest and a similar approach can be used to  to prove the correctness of other classical distributed building blocks (\emph{e.g.} agreement, clock synchronization, interactive consistency etc)  under Mobile Byzantine Faults model.

%% file: relatedWork.tex
%
%

The Byzantine Agreement problem, introduced first by Lamport \emph{et al.}~\cite{LSP82} is one of the most studied building blocks in distributed computing and is specified as the conjunction of the following three properties~\cite{Lynch96}: \emph{(Termination)} All correct processes eventually decide; \emph{(Agreement)} No two correct processes decide on different values; \emph{(Validity)} If all correct processes start with the same value $v$, then $v$ is the only possible decision value for a correct process.

In this paper we are interested in the Approximate Byzantine Agreement where processes start with real numbers as inputs, and eventually decide a real number as output. The difference with the (exact) Byzantine Agreement is that instead of agreeing exactly, processes are allowed to disagree within a small positive margin $\epsilon$. The specification of the Approximate Byzantine Agreement \cite{Lynch96} has the same termination property as the Byzantine Agreement. However, it has different agreement and validity properties: \emph{($\epsilon$-Agreement)} for any $\epsilon>0$, the decision values of any pair of correct processes are within $\epsilon$ of each other; \emph{(Validity)} any decision value for a correct process is in the range of the initial values of the correct processes.

\subsection{Approximate Byzantine Agreement}

The Approximate Byzantine Agreement problem has been studied since the eighties~\cite{DLPS86,FLM86}. Most of the presented solutions are based on successive rounds of exchanges of the latest values each process stores locally. Upon collecting each set of values, a correct process applies a function (\emph{e.g.} average) and adopts as next value the value returned by the function. The interested reader may refer to reference textbooks~\cite{Lynch96} and references herein~\cite{Fekete90,Fekete94}.

Allowing different kinds of faults was investigated by Kieckhafer \emph{et al.}~\cite{KA94}, as they unify different algorithms into the class of MSR-algorithms (Mean-Subsequence-Reduced), which compute the mean of a subsequence of the reduced multi-set of values. The authors analyze the convergence rate and the fault-tolerance of this class of algorithm in a so-called \emph{Mixed-Mode faults model}. In this model faults are partitioned into asymmetric (classical Byzantine), symmetric and benign. The benign faults are self-incriminating (immediately self-evident to all non faulty processes). The behavior of symmetric faults is perceived identically to all correct processes, while the asymmetric faults have a totally arbitrary behavior. That is, the behavior of processes being subject to asymmetric faults may be perceived differently by different correct processes. 

Stolz \emph{et al.}~\cite{SW15} recently proposed an Approximate Byzantine Agreement solution where processes have to approximate the median value of the input values. Their algorithm achieves agreement for $n>3f+1$ within $f+1$ rounds, where $f$ denotes the number of faulty (\emph{a.k.a.} Byzantine) processes, while $n$ denotes the total number of processes. Their algorithm is not included in the MSR-class of \cite{KA94} since they use a variant of the King algorithm~\cite{BGP89}. Multidimensional  agreement has been investigated by Mendes \emph{et al.}~\cite{MH13,MHVG15}, where the authors also highlight the connexion between approximate agreement and convergence in mobile autonomous robot networks~\cite{BPT09c,BPT10j}. Li \emph{et al.}~\cite{LHW14} and Charron-Bost \emph{et al.}~\cite{Charron-BostFN15} consider extensions to dynamic networks.      
In a sustained line of work, Tseng \emph{et al.}~\cite{SV15,TV13,TV14a,TV14b,VTL12} investigate approximate agreement problem within various faults models (link crash, process crash, byzantine) in multi-hop networks (both for the directed and the undirected cases).
        


\subsection{Mobile Byzantine Faults}

As singled out by Yung~\cite{Y15}, it is worth considering \emph{mobile adversaries (a.k.a. Byzantine mobile agents)}. Mobile adversaries have been primarily introduced in the context of multi-party computation, to model an attacker or an adversarial environment that is able to progressively compromise computational entities, but only for a limited period of time. Therefore, tolerating Mobile Byzantine Faults is, in some sense, like having a bounded number of compromised entities at any given time but the set of such entities evolves over time. 
 
From a theoretical point of view, mobile adversaries have been formalized for different \emph{Mobile Byzantine Faults} models \cite{BDNP14,Garay+95+ORA,Garay+1994,Sasaki+2013}.
In Mobile Byzantine Faults models, there are two main research directions: \emph{(i)} Byzantines with constrained mobility, and \emph{(ii)} Byzantines with unconstrained mobility.  Byzantines with constraint mobility were first studied by Buhrman \emph{et al.} \cite{Garay+95+ORA}. In their paper, they consider that Byzantine agents move from one node to another only when protocol messages are sent (similarly to how viruses would propagate). 
Buhrman \emph{et al.} \cite{Garay+95+ORA} studied the problem of Mobile Byzantine Agreement. They proved a tight bound for the problem solvability (\emph{i.e.}, $n > 3t$, where $t$ is the maximal number of simultaneously faulty processes), and proposed a time optimal protocol that matches this bound. 

In the case of unconstrained mobility the motion of Byzantine agents is not tied to protocol message exchanges. Several authors investigated the agreement problem in further variants of this model \cite{Banu+2012,BDNP14,Garay+1994,Ostrovsky+91,Reischuk+85,Sasaki+2013}.
Reischuk \cite{Reischuk+85} investigates the stability/stationarity of malicious agents for a given period of time.  Ostrovsky and Yung \cite{Ostrovsky+91} introduce the notion of mobile virus and investigate an adversary that can inject and distribute faults. Furthermore, they advocate that the unconstraint mobility model abstracts the concept of insider threats (hacker, cracker, black hat) or attacks (DOS, Worms, viruses or Trojan horses). 
Garay \cite{Garay+1994} and, more recently, Banu \emph{et al.} \cite{Banu+2012}, Sasaki  \emph{et al.}  \cite{Sasaki+2013}, and Bonnet  \emph{et al.} \cite{BDNP14} consider, in their models, that processes execute synchronous rounds composed of three phases: \emph{send}, \emph{receive}, \emph{compute}. Between two consecutive rounds, Byzantine agents can move from one node to another, hence the set of faulty processes has a bounded size although its members can change from one round to the next. 
%
The main difference between the aforementioned unconstrained models lies in the knowledge that processes that have been affected by a Byzantine agent have. In Garay's model, a process has the ability to detect its own infection after the Byzantine agent left it. More precisely, during the first round following the leave of the Byzantine agent, a process enters a state, called \emph{cured}, during which it can take preventive actions to avoid sending messages that are based on a corrupted state. Garay \cite{Garay+1994} proposed, in this model, an algorithm that solves Byzantine Agreement provided that $n>6f$ (this requirement was later dropped to $n>4f$ \cite{Banu+2012}). 
Bonnet  \emph{et al.} \cite{BDNP14}  investigated the same problem in a  model where processes do not have the ability to detect when Byzantine agents have moved. However, differently from Sasaki  \emph{et al.}  \cite{Sasaki+2013}, cured processes have \emph{control} on the messages they send. This subtle difference on the power of Byzantine agents has an impact on the bounds for solving the agreement. If in the Sasaki's model the bound on solving agreement is $n>6f$,  in Bonnet's model it decreases to $n>5f$, and this bound is proven tight.

%% file: contributions.tex

%
%

%% file: systemModel.tex

We consider a distributed system composed of a set of $n$ processes $\mathcal{P}=\{p_1, p_2, \dots p_n\}$ each having a unique integer identifier $i \in [1, n]$.\\
 
\noindent {\bf Communication model and timing assumptions.} Processes communicate through message passing. It is assumed that processes in the distributed system may access a built-in communication abstraction used to disseminate messages to all the other processes. We assume that communications  are authenticated (\emph{i.e.}, given a message $m$, the identity of its sender cannot be forged) and reliable (\emph{i.e.} messages are not created, lost or duplicated).

The system is synchronous and evolves in sequential synchronous rounds $r_0, r_1,$ $\dots r_i \dots$. Every round is divided in three phases: (i) \emph{send} where processes send all the messages for the current round, (ii) \emph{receive} where processes receive all the messages sent at the beginning of the current round\footnote{Let us note that, in round-based computations, all messages are delivered during the receive phase.} and (iii) \emph{computation} where processes process received messages and prepare those that are sent in the next round. 

\noindent{\bf Failure model.} Processes are affected by \emph{mobile Byzantine failures} (MBF) \cite{BDNP14,Garay+1994,Garay+95+ORA,Sasaki+2013}.
Informally, in the mobile Byzantine failure model, faults are represented by powerful computationally unbounded agents that move arbitrarily from a process to another. When the agent is on the process, it can corrupt its local variables, forces it to send arbitrary messages (potentially different from process to process) etc... However, the agent cannot corrupt the identity of the process.
We assume that, in each round $r_i$, at most $f$ processes can be affected by a mobile Byzantine failure. 
When an agent occupies a process $p_i$ we say that $p_i$ is {\em faulty}. If a process has been occupied by a Byzantine agent in the previous round then the process is said to be {\em cured}. If a process is neither {\em faulty} nor {\em cured} then it is said to be {\em correct}.  
We assume, similar to previous work \cite{BDNP14,Garay+1994,Sasaki+2013}, that each process has a tamper-proof memory where it safely stores the correct algorithm code.
When the agent leaves a process $p_i$, it becomes {\em cured} and then can recover the correct algorithm code from the tamper-proof memory.
Concerning the assumptions on agent movements and the process awareness on its {\em cured} state, different models have been defined. In this paper we consider all the variants of mobile Byzantine failures \cite{BDNP14,Garay+1994,Garay+95+ORA,Sasaki+2013}:

\begin{itemize}
	\item {\bf (M1)} \emph{Garay's model} \cite{Garay+1994}. In this model, agents can move arbitrarily from a process to another at the beginning of each round (\emph{i.e.} before the send phase starts). When a process is in the {\em cured} state it is aware of its condition and thus can remain silent for a round to prevent the dissemination of wrong information. 
	\item {\bf (M2)} \emph{Bonnet et al.'s model} \cite{BDNP14} and {\bf (M3)} \emph{Sasaki et al.'s model} \cite{Sasaki+2013}. As in the previous model, agents can move arbitrarily from a process to another at the beginning of each round (\emph{i.e.} before the send phase starts). Differently from the Garay's model, in both models it is assumed that processes do not know if they are correct or cured when the Byzantine agent moved. The main difference between these two models is that in the \cite{Sasaki+2013} model a cured process still acts as a Byzantine one extra round.  
	\item {\bf (M4)} \emph{Buhrman's model} \cite{Garay+95+ORA}. Differently from the previous models, agents move together with the message (\emph{i.e.}, with the ${\sf send}$ or ${\sf broadcast}$ operation). However, when a process is in the {\em cured} state it is aware of that. 
\end{itemize}

\noindent {\bf Byzantine Approximate Agreement specification.} 
The Byzantine Approximate Agreement problem has been accurately specified in \cite{Lynch96}.  Processes start with real-valued inputs and eventually decide a real-valued output. 
The only difference with the exact Byzantine Agreement is that instead of agreeing exactly, processes are allowed to disagree within a small positive real-valued tolerance $\epsilon$. 
\begin{itemize}
\item{Termination}: All non faulty processes eventually decide;
\item{$\epsilon$-Agreement}:  The decision value of any pair of non faulty processes are within $\epsilon$ of each other;
\item{Validity}:  Any decision value for a non faulty process is in the range of the initial values of the non faulty processes.
\end{itemize} 
Note that the specification proposed in \cite{LSP82} is similar (the termination properties being included in the agreement properties).

%% file: background.tex

In this paper, we extend the analysis done in \cite{KA94} for mixed-faults model and prove that the family of \emph{Mean-Subsequence-Reduce} (MRS) algorithms works also in the Mobile Byzantine Faults models.

In this section, we provide some background notions from \cite{KA94} and propose an elegant mapping between the Mobile Byzantine Faults model and the Mixed-Mode faults.

The work in \cite{KA94} is focused on a specific family of Byzantine Approximate Agreement algorithms, namely \emph{convergent voting algorithms}, that start from an initial set of proposed values $\{v_1, v_2, \dots v_n\}$ and guarantee that any process $p_i$ 
converges to a value $v_i$ satisfying the  Byzantine Approximate Agreement specification.
More in details, any algorithm in this family proceeds in rounds and during any round $r_j$, every process $p_i$ executes the following actions:

\begin{enumerate}
\item \emph{send-phase}: $p_i$ sends its \vir{voted} value to the others; 
\item \emph{received-phase}: $p_i$ aggregates values in a multiset $N_{r_k}$;  
\item \emph{computation-phase}: $p_i$ applies a deterministic function $\mathcal{F}(N_{r_k})$ to decide the value to vote in the next round $r_{k+1}$.
\end{enumerate}

In 
 \cite{KA94} \emph{convergent voting algorithms}  are called \emph{Mean-Subsequence-Reduce} (MSR). Their computation function can be expressed in the general form: 

$$\mathcal{F}_{MSR}(N_{r_k})= {\sf mean}[Sel(Red(N_{r_k}))]$$ where $Sel$ is a selection function and $Red$ is a reduction function used to filter values. 

The correctness of MSR algorithms in the Mixed-mode faults model is guaranteed by the \emph{single-step convergence} property. 
Informally, at the end of each round $r_k$, the range of values voted by correct processes shrinks with respect to the beginning of the round. 
The failures considered in \cite{KA94} are benign, symmetric and asymmetric with the definitions below.

\begin{definition}[Benign fault \cite{KA94}] 
A process $p_i$ is said to be benign faulty if it exposes a self-incriminating, or immediately self-evident fault to all non-faulty processes.
\end{definition}

An example of benign fault is a crash failure or an omitted reply in a synchronous system. That is, in synchronous systems  if the reply is not delivered within the expected time then the process can be immediately detected as faulty by every correct process .

\begin{definition}[Symmetric fault \cite{KA94}]
A process $p_i$ is said to be symmetrically faulty if its behavior is perceived identically by all non-faulty processes.
\end{definition}

A symmetric fault is generally a malicious fault such as unexpected message broadcast to all processes.

\begin{definition}[Asymmetric fault \cite{KA94}]
A process $p_i$ is said to be asymmetrically faulty	 if its behavior may be perceived differently by different non-faulty processes. 
\end{definition}

An asymmetric fault is a classical arbitrary fault such as a broadcast where the sender can send different values to different correct processes.

In \cite{KA94}, the authors proved that, given the number of benign faults $b$, the number of symmetric faults $s$ and the number of asymmetric faults $a$, the minimum number of processes $n$ needed to solve the Byzantine Approximate Agreement by an algorithm in the class MSR is $n> 3a+2s+b$.


In the following, we propose a method to map the Mobile Byzantine faults model to the Mixed-mode faults then prove that the MSR algorithms are correct under the Mobile Byzantine fault model. 
In addition, we compute the number of processes $n$ needed to tolerate $f$ Mobile Byzantine faulty processes  and solve the Byzantine Approximate Agreement problem under Mobile Byzantine faults model.



%

%
Note that the behavior of mobile Byzantines concerns only the send/receive phases of the MSR algorithms. Therefore, we focus on the behavior of the faulty processes during the execution of these phases.
In order to match our models the send-phase of MSR algorithms should be sightly modified in order to prevent correct processes to participate to the communication as per the requirement of the {\bf M1} model.
 
\begin{lemma}\label{l:mappingb}
Let $\mathcal{T}b_{r_k}$ be the set of cured processes at the beginning of round $r_k$ in model {\bf M1}. 
If the  send phase
$${\bf if}~(cured)~\emph{nop}; ~{\bf else}~ send(vote)~ to~ all~ processes;$$
is executed by any $p_j \in \mathcal{T}b_{r_k}$ then the computation executed in round $r_k$  is equivalent to the computation under Mixed-mode fault model with $a=f$ and $b=|\mathcal{T}b_{r_k}|$.
\end{lemma}

\begin{proofL}
A cured process, in {\bf M1} is aware of its failure state thus if it is forced to skip the send phase then it is detected by any correct process in round $r_k$.
	\renewcommand{\toto}{l:mappingb} 
\end{proofL}

\begin{lemma}\label{l:mappings}
Let $\mathcal{T}s_{r_k}$ be the set of cured processes at the beginning of round $r_k$ in model {\bf M2}. 
	If the send phase 
	$$send(vote)~ to~ all~ processes;$$
	is executed by any $p_j \in \mathcal{T}s_{r_k}$ then the computation executed in round $r_k$ is equivalent to the computation under Mixed-mode fault model executed with $a=f$ and $s=|\mathcal{T}s_{r_k}|$.
\end{lemma}

\begin{proofL}
	A cured process in {\bf M2} is not aware of its state, hence it sends its vote to every process in the system. This value may be the result of a corrupted state.
	This is identical to the behavior of  a process exhibiting a symmetric fault.
	\renewcommand{\toto}{l:mappinga}
\end{proofL}

\begin{lemma}\label{l:mappinga}
Let $\mathcal{T}a_{r_k}$ be the set of cured processes at the beginning of round $r_k$ in model {\bf M3}. 
	If send phase 
	$$send(vote)~ to~ all~ processes;$$
	is executed by any $p_j \in \mathcal{T}a_{r_k}$ then the computation executed in round $r_k$ is equivalent to the computation under Mixed-mode fault model executed with $a=f+|\mathcal{T}a_{r_k}|$.\\
\end{lemma}
\begin{proofL}
	A cured process in {\bf M3} is not aware of its state hence it sends its vote to every process in the system. Moreover, Byzantine agent prepares the outgoing message queue (cf. \cite{Sasaki+2013}). Thus, a cured process executes the sending phase as any correct process. However, differently from the correct processes it sends possibly different values (left behind by the Byzantine agent) to every process in the system. This is identical to the behavior of a process exhibiting  an asymmetric fault.
	\renewcommand{\toto}{l:mappinga}
\end{proofL}

\begin{lemma}\label{l:mappingn}
	Let $\mathcal{T}c_{r_k}$ be the set of cured processes at the beginning of round $r_k$ in model {\bf M4}.
	If the send phase
	$$send(vote)~ to~ all~ processes;$$
	is executed by any $p_j \in \mathcal{T}c_{r_k}$ then the computation executed in round $r_k$ is equivalent to the computation under Mixed-mode fault model executed with $a=f$. 
\end{lemma}

\begin{proofL}
	In this failure model, Byzantine agents move along with the  messages. Thus during the sending phase there are no processes in $\mathcal{T}c_{r_k}$. 
	\renewcommand{\toto}{l:mappingn}
\end{proofL}

Table \ref{tab:mapping} summarizes the mapping results proven in Lemmas \ref{l:mappingb}-\ref{l:mappingn}. Table \ref{tab:mapping2} reports the required number of replicas for each model.
\begin{table}[]
	\centering
	\begin{tabular}{|l|c|c|c|c|}
		\hline
		 & {\bf M1} & {\bf M2} & {\bf M3} & {\bf M4} \\ \hline
		{Asymmetric}  & faulty & faulty & faulty, cured& faulty \\ \hline
		{Symmetric}& & cured  & &  \\ \hline
		{Benign} & cured & & &  \\ \hline
	\end{tabular}
	\caption{Mapping between the behavior of faulty processes in the Mixed-Mode faulty model and faulty and cured processes in the four Mobile Byzantine faulty models.}
	\label{tab:mapping}
\end{table}

\begin{table}[]
	\centering
	\begin{tabular}{|l|c|}
		\hline
		& $n_{Mi}$\\ \hline
		{\bf M1} & $n > 3f+b=4f$ \\ \hline
		{\bf M2} & $n > 3f+2s=5f$ \\ \hline
		{\bf M3} & $n > 3(f+a)=6f$ \\ \hline
		{\bf M4} & $n >3f=3f$ \\ \hline

	\end{tabular}
	\caption{Number of required replicas in each failure model.}
	\label{tab:mapping2}
\end{table}

%% file: mappingProofs.tex
In the following we prove that in presence of mobile Byzantine agents the MSR family of algorithms verifies the Byzantine Approximate Agreement specification. We first characterize configurations produced by a MSR algorithm in presence of static Byzantine faulty nodes. Then, we prove that each configuration produced in  presence of mobile Byzantine agents has the same characterization. Hence, the mobility of Byzantine agents does not affect the correctness of MSR family. 
Moreover, we prove that the necessary condition over the number of replicas in \cite{KA94} still holds in the Mobile Byzantine failures model with the mapping defined in the previous section. 

\subsection{Preliminaries}\label{ss:preliminaries} In the following we recall some definitions from \cite{DLPS86, KA94} :
\begin{itemize}
	\item{min(V):} min($r \in \mathbb{R}: V(r)>0) = v_1$; the minimum value of the elements in $V$;
	\item{max(V):} max($r \in \mathbb{R}: V(r)>0) = v_v$; the maximum value of the elements in $V$;
	\item{$\rho(V)$:} $[min(V),max(V)] = [v_1,v_v]$; the real interval spanned by V. $\rho(V)$ is called the range of V;
	\item{$\delta(V)$:} $min(V)-max(V) = v_1 - v_v$; the difference between the maximum and the minimum values of $V$. $\delta(V)$ is called the diameter of V;
	\item{$N_{r_k}^i$:} the multiset of values received in a given round $r_k$ by non-faulty process $i$. Let  {$U$:} be the subset of  $N_{r_k}^i$, the values generated by non-faulty processes \footnote{Since the communication graph is fully connected then this set is equal for any correct process}.
\end{itemize}

Now we can recall the important properties of $\mathcal{F}_{MSR}()$ as proved in \cite{KA94}.
If $n>3a+2s+b$  then the following two properties hold:
	\begin{itemize}
		\item[P1] For each non faulty process $p_i$, the computed value is in the range of non faulty values, i.e., $\mathcal{F}_{MSR}(N_{r_k}^i) \in \rho(U)$.
		\item[P2] For each pair of non faulty processes, $p_i$ and $p_j$, the difference between their computed values is strictly less than the diameter of the submultiset of non faulty values received, i.e., $|\mathcal{F}_{MSR}(N_{r_k}^i)-\mathcal{F}_{MSR}(N_{r_k}^j)|< \delta(U)$.
	\end{itemize}
In the following $v^i_{r_k}$ denotes the value obtained at the end of round $r_k$ (computation phase) by process $p_i$, applying the MSR function vector $N_{r_k}^i$.

\begin{definition}[correct value]
	Given a value $v^i_{r_k} \leftarrow \mathcal{F}_{MSR}(N_{r_k}^i)$, $v^i_{r_k}$ is said to be correct if it respects the two  $\mathcal{F}_{MSR}()$ function properties $P1$ and $P2$.
\end{definition}

\begin{lemma}\label{l:curedOnRound}
	Let $\mathcal{T}*_{r_k}$ be the set of cured processes at the beginning of round $r_k$ in the models {\bf M1-M4}. If $n>n_{Mi}$  and every $p_j \in \mathcal{T}*_{r_k}$ executes computation-phase of a MSR-algorithm then at the end of $r_k$ we have $|\mathcal{T}*_{r_k}|=0$.
\end{lemma}

\begin{proofL}
	The proof is done by induction. 
	During the first round $r_0$ no Byzantine agent moved yet. Thus, at the end of $r_0$ trivially $|\mathcal{T}*_{r_0}|=0$. 
	In the next round $r_1$ Byzantine agents move thus affecting up to $f$ processes. Therefore, at the beginning of $r_1$ there are up to $f$ cured processes, $|\mathcal{T}*_{r_1}|\leq f$. If we substitute, for each model {\bf M1-M4} (cf. Table \ref{tab:mapping}), values in $n>3a+2s+b$ if follows that despite agents movement, $n>n_{M_i}$ still holds. Thus, for the definition of $\mathcal{F}_{MSR}()$ the value that each process computes at \emph{computation-phase} is correct. Hence, at the end of round $r_1$ we have $|\mathcal{T}*_{r_1}|=0$.
	For each further $r_k$ the reasoning is similar.
	\renewcommand{\toto}{l:curedOnRound}
\end{proofL}

From Lemma \ref{l:curedOnRound} it follows that during each round there are not cured processes related to the previous round but only the ones due to the last Byzantine agents movement, hence the corollary below. 

\begin{corollary}\label{c:numberOfCured}
	Let $\mathcal{T}_{r_k}$ be the set of cured processes at the beginning of round $r_k$. $\forall r_k$, $|\mathcal{T}_{r_k}|\le f$.
\end{corollary}


\begin{definition}[configuration $C_{r_k}$]
Let configuration $C_{r_k}$ be a set of $n$ tuples $\langle$failure state, proposing value$\rangle_i$ representing the state of each process $p_i$ at round $r_k$. Note that processes, depending on the failure model, may or may not be aware of their failure state.
\end{definition}

%
%
%

\begin{definition}[$AA_{r_k}$]
Let  $\mathcal{AA}$ be a generic instance of the MSR family  and let $AA_{r_k}$ be the $r_k-th$ execution of the protocol $\mathcal{AA}$ at round $r_k$, such that $C_{r_k} \leftarrow AA_{r_k}(C_{r_k-1})$. It takes as input $C_{r_k-1}$ and returns $C_{r_k}$.
\end{definition}

\begin{definition}[static computation]
A sequence of $k$ $\mathcal{AA}$ executions, such that $C_{r_k}\leftarrow AA_{r_k-1}(AA_{r_k-2}(\dots$ $ AA_{r_1}(C_{r_0})) \dots)$ is said a static computation if in every configuration $C_{r_1},...,C_{r_k}$, there exists a subset of at least $n-(3a+2s+b)$ correct processes that are correct during the whole computation. 
\end{definition}

Note that  with fixed $a$,$s$ and $b$, the relation $n>3a+2s+b$ always holds in a static computation of a MSR algorithm (\cite{KA94}). 

\begin{definition}[mobile computation]
A sequence of $k$ $\mathcal{AA}$ executions, such that $C_{r_k}\leftarrow AA_{r_k-1}(AA_{r_k-2}(\dots$ $ AA_{r_1}(C_{r_0})) \dots)$ is said to be a mobile computation if for any two subsequent configurations $C_{r_k}$, $C_{r_k+1}$, any process may change the failure state but the relation $n>3a+2s+b$ holds at each round. 
\end{definition}

\begin{definition}[configurations equivalence]
A configuration $C_{r_k}$ is said to be equivalent to a configuration $\bar{C}_{r_k}$ if:
\begin{itemize}
\item $C_{r_k}$ and $\bar{C}_{r_k}$
 produce the same $U$;
\item $\forall k$, $C_{r_k}$ has at least the same number of tuples  $\langle$correct, correct value $\rangle$ as $\bar{C}_{r_k}$.
\end{itemize}
Note that in a static computation a correct process is correct for the whole computation, while in a mobile one is correct with respect to the observed round. 
\end{definition}

\begin{definition}[correct computation]
A computation $C_{r_0}, \dots, C_{r_k}$ is a correct computation if it is possible to build a static computation $\bar{C}_{r_0}, \dots, \bar{C}_{r_k}$ such that, $\forall j \in[0,k]$, $C_{r_j}$ is equivalent to $\bar{C}_{r_j}$. 
\end{definition}

\begin{observation}\label{o:correctConf}\cite{KA94}
Given a static computation $\bar{C}_{r_0}, \dots, \bar{C}_{r_k}$ of an algorithm in the MSR class, if $n>3a+2s+b$, then 
each configuration $\bar{C}_{r_j}, j \in [0,k]$, is characterized as follows:
\begin{itemize}
\item up to $a$ asymmetric Byzantine processes;
\item up to $s$ symmetric Byzantine processes;
\item up to $b$ benign faults;
\item at least $n-(a+s+b)$ correct processes such that  each $p_j$ of them computes a correct value $v^{r_j}_j$.
\end{itemize}
\end{observation}

The first three points are due to the failures static nature. The last one is given by the failures static nature plus the correctness of the algorithm in the static case (as proven in \cite{KA94}).

\subsection{MSR correctness under Mobile Byzantine fault model}
In the following we prove that despite Byzantines mobility, the MSR family of algorithms verifies the Approximate Agreement specification. In the presence of mobile Byzantine agents, each round is characterized by correct, cured and faulty processes. As we showed previously, depending on the failure model considered, cured processes behave accordingly to a different kind of fault (asymmetric, symmetric or benign). 


%
The following theorem proves the mapping between the Mobile Byzantine faults model and the Mixed-mode fault model. Let us start proving that if $n>n_{Mi}$ then a mobile computation is also a correct computation, as defined in subsection \ref{ss:preliminaries}.
\begin{theorem}\label{l:mapping}
Let us consider a mobile computation $C_0, \dots, C_k, \forall k \in \mathbb{N}$ of an algorithm  $\mathcal{AA}$ in the class MSR. If in each round $n>n_{Mi}$ (cf. Table \ref{tab:mapping2}) then the sequence $C_0, \dots, C_{k}$ is a correct computation.
\end{theorem}

\begin{proofT}
We have to show that for each iteration of $\mathcal{AA}$ we can build a static computation equivalent to the dynamic one. The proof is done by induction. Let us denote by $\mathcal{C}$, $\mathcal{T}*$ and $\mathcal{B}$ the set of correct, cured and Byzantine processes respectively and let $t_*$ denote the cardinality of $\mathcal{T}*$. Let us denote, in the static case, by $\mathcal{C}'$, $\mathcal{T}'$, and $\mathcal{B}'$ the set of correct, non correct (which may be asymmetric, symmetric, or benign), and asymmetric faulty processes, respectively, and let $t'_*$ denote the cardinality of $\mathcal{T}'$.

\begin{itemize}
\item{Rounds $0 \rightarrow 1$:} At the begining of round 0, Byzantine agents never move. Thus, the configuration is as follows: 
\begin{itemize} 
\item{$\mathcal{C}$}: $\forall i \in \mathcal{C}, \langle correct, v_i^{init} \rangle_i$, $|\mathcal{C}|\geq n-(f)$;
\item{$\mathcal{B}$}: $\forall j \in \mathcal{B},\langle faulty, \bot$ \footnote{We use $\bot$ to indicate that it can be any value} $\rangle_j$, $|\mathcal{B}| \leq f$.
\end{itemize}
The protocol executes its first iteration. 
Processes exchange their value and each non Byzantine process $p_i$ updates its state: $\langle$failure state, proposing value $\leftarrow v_i^0=\mathcal{F}_{MSR}(V^{0})\rangle$ . 
At this point the situation is as follow:
\begin{itemize}
\item{$\mathcal{C}$}: $\forall i \in \mathcal{C}, \langle correct, v_i^{0} \rangle_i$, $|\mathcal{C}|\geq n-(f)$;
\item{$\mathcal{B}$}: $\forall j \in \mathcal{B},\langle faulty, \bot \rangle_j$, $|\mathcal{B}| \leq f$.
\end{itemize}
Up to now, the same happens in a static computation. 
At the begining of round 1, at most $f$ Byzantine agents move affecting other processes. Thus there are up to $t_*=f$ cured processes storing a non correct value (e.g., $v^0 \notin \rho(N^{0})$). 
\begin{itemize}
\item{$\mathcal{C}$}: $\forall i \in \mathcal{C}, \langle correct, v_i^{init} \rangle_i$, $|\mathcal{C}|\geq n-(f+t_*)$;
\item{$\mathcal{T}$}: $\forall k \in \mathcal{T}, \langle cured, \bot \rangle_k$, $|\mathcal{T}|\leq t_*$;
\item{$\mathcal{B}$}: $\forall j \in \mathcal{B},\langle faulty, \bot \rangle_j$, $|\mathcal{B}| \leq f$.
\end{itemize}
At the begining of round 1, there are at least $n-(f+t_*)$ correct processes. If we map it to the Mixed-mode failures model (cf. Table \ref{tab:mapping}), this is equivalent to a static configuration where there are $f$ asymmetric processes and $t_*$ non correct that may be asymmetric, symmetric or benign:
\begin{itemize}
\item{$\mathcal{C}'$}: $\forall i \in \mathcal{C}', \langle correct, v_i^{init} \rangle_i$, $|\mathcal{C}'|\geq n-(f+t'_*)$;
\item{$\mathcal{T}'$}: $\forall k \in \mathcal{T}', \langle *, \bot \rangle_k$, $|\mathcal{T}'|\leq t'_*$;
\item{$\mathcal{B}'$}: $\forall j \in \mathcal{B}',\langle asymmetric, \bot \rangle_j$, $|\mathcal{B}'| \leq f$.
\end{itemize}
The mobile and static configurations are equivalent (cf. Observation \ref{o:correctConf}). Thus the current mobile configuration (and the mobile computation up to now) is correct.

\item{Rounds $1 \rightarrow 2$:} From the previous point, the configuration at the beginning of round 1 is correct. The second iteration of the protocol takes place. Processes exchange their value and each non Byzantine process $p_i$ updates its state: $\langle$failure state, proposing value $\leftarrow v^1_i=\mathcal{F}_{MSR}(N^{1}_i)\rangle$. At this point, for Lemma \ref{l:curedOnRound}, each process in $\mathcal{T}*$ becomes correct. In other words, there are up to $f$ Byzantine processes and at least $n-f$ correct processes. We are in the same situation as at the end of previous round $0$.\\
At the beginning of next round, at most $f$ Byzantine agents can move to other processes, leaving up to $t_*=f$ cured processes with non correct value. 
Thus there are at least $n-(f+t_*)$ correct processes at the begining of round 2. 
The mobile and static configurations are equivalent (cf. Observation \ref{o:correctConf}). Thus the current mobile configuration (and the mobile computation up to now) is correct.
\item{Rounds $i \rightarrow i+1$:} generalizing, for each round starting with a correct configuration we can apply the previous reasoning ending in a subsequent round characterized by a correct configuration.
\end{itemize}
\renewcommand{\toto}{l:mapping}
\end{proofT}

%
In the following we prove the correctness of any algorithm in the class MSR under Mobile Byzantine failure model.
\begin{lemma}[Termination]\label{l:termination}
	Let $AA$ be an algorithm in the class MSR.   If $n>n_{Mi}$, $AA$ under Mobile Byzantine fault model verifies the \emph{Termination} property of the Byzantine Approximation Agreement.
\end{lemma}

\begin{proofL}
From Theorem \ref{l:mapping}, if $n>n_{Mi}$ then algorithm $AA$ generates a sequence of correct configurations, i.e., a sequence of converging values exactly as in \cite{DLPS86,KA94}, thus the Termination property is satisfied in the same way this is satisfied by the \cite{DLPS86,KA94} solutions.	
	\renewcommand{\toto}{l:termination}
\end{proofL}

\begin{lemma}[$\epsilon$-Agreement]\label{l:agreement}
Let $AA$ be an algorithm in the class MSR.   If $n>n_{Mi}$, $AA$ under Mobile Byzantine fault model verifies the \emph{$\epsilon$-Agreement} property of the Byzantine Approximation Agreement.
%
\end{lemma}

\begin{proofL}
From Theorem \ref{l:mapping}, if $n>n_{Mi}$ then algorithm $AA$ generates a sequence of correct configurations, i.e., a sequence of converging values exactly as in \cite{DLPS86,KA94}. Thus, the $\epsilon$-Agreement property is satisfied in the same way this is satisfied by the \cite{DLPS86,KA94} solutions. 

In the following we prove that once $\epsilon$-Agreement is achieved among the currently non faulty processors, it is preserved among the (possible different) uninfected processors.
	Let us consider an arbitrarily long mobile computation $C_0, \dots, C_{k}$. If $\epsilon$-Agreement is achieved then there exists a round $r_a, a \in [0,k]$ where all non faulty processes agree on values that are $\epsilon$ close to each other. Considering that $n>n_{Mi}$ then from Theorem \ref{l:mapping} the whole mobile computation $C_0, \dots, C_{k}$ is correct. Thus from round to round the two properties $P1$ and $P2$ hold and correct processes values can not diverge from each other.

\renewcommand{\toto}{l:agreement}
\end{proofL}

\begin{lemma}[Validity]\label{l:validity}
Let $AA$ be an algorithm in the class MSR.   If $n>n_{Mi}$, $AA$ under Mobile Byzantine fault model verifies the \emph{Validity} property of the Byzantine Approximation Agreement.
\end{lemma}

\begin{proofL}
	From Theorem \ref{l:mapping}, if $n>n_{Mi}$ then algorithm $AA$ generates a sequence of correct configurations, i.e., a sequence of converging values exactly as in the validity proof in \cite{DLPS86,KA94}. 
	\renewcommand{\toto}{l:validity}
\end{proofL}

The three above lemmas provide the proof of the theorem below.
\begin{theorem}\label{t:mapping}
If $n>n_{M_i}$ then the class  MSR verifies the Byzantine Approximate Agreement specification.
\end{theorem}
%

%% file: lowerBounds.tex
In order to formulate the strongest impossibility results related to Approximate Agreement in the Mobile Byzantine faults model we examine a  weaker version of this problem referred in  \cite{FLM86}
as \emph{Simple Approximate Agreement}. Each correct node has a real value from $[0,1]$ as input and chooses a real value. Correct behaviors must satisfy the following properties: 
\emph{Agreement:} The maximum difference between values chosen by correct nodes must be strictly smaller than the maximum difference between the inputs, or be equal to the latter difference if it is zero. \emph{Validity}: Each correct node chooses a value in the range of the inputs of the nodes.

%
We prove lower bounds for each Mobile Byzantine faults models: Garay\rq{}s (M1), Bonnet\rq{}s(M2), Sasaki\rq{}s (M3) and Burhman\rq{}s (M4). The bounds for the models (M3) and (M4) result from the classical bounds proved in \cite{FLM86} and the mapping defined in Section \ref{section:systemmodel}. In the case of models (M1) and (M2), since the behavior of cured processes cannot be totally controlled by the Byzantine adversary, specific proofs are needed.
\begin{observation}
Note that the lower bounds below do not concern the class of algorithms whose computations end before the end of the first round and that start in a configuration where there are $f$ Byzantine processes and no {\em cured} ones. It is trivial that for this class of algorithms the lower bounds are the same as those proven in \cite{FLM86} (i.e., $n \geq 3f+1$).
\end{observation}

 

\begin{theorem}[Lower bound for Garay\rq{}s model]
\label{t:garayBound}
There is no algorithm that solves Simple Approximate Agreement in the Garay's model (M1) under the Mobile Byzantine faults model  if $n \leq 4f$.
\end{theorem}

\begin{proofT}
The proof goes by contradiction.
Suppose that  there exists an algorithm $\mathcal{A}$ verifying the Simple Approximate Agreement properties in the (M1) Mobile Byzantine faults model with $n \leq 4f$. 
Consider w.l.g. a system with four processes and one Byzantine mobile agent. The generalization of the proof can be done by replacing any process with a group of  $f$ processes.

%
Consider the system with four processes denoted $p_0,p_1,p_2,p_3$ and consider that  $p0$ is occupied by the Byzantine agent while $p_1$ is cured and $p_2$ and $p_3$ are correct processes. Note that the cured process in (M1) model is silent.
Consider three executions of $\mathcal{A}$ denoted $E1$, $E2$ and $E3$ constructed as follows.
In $E1$ the correct processes propose both the value $0$. It follows,  from the Agreement and Validy properties of $\mathcal{A}$, that the value chosen by $p_1,p_2$ and $p_3$  should be $0$ (independently of the value sent by the Byzantine process, assume it $1$).  In $E2$ the correct processes propose both $1$. It follows, from the Agreement and Validity properties of $\mathcal{A}$, that the value chosen by $p_1,p_2$ and $p_3$ is $1$ (independently of the value sent by the Byzantine process, assume it $0$). 

The $E3$ brings the contradiction: some correct processes choose $1$ while others choose $0$, which contradicts the Agreement property of $\mathcal{A}$. The execution $E3$ is as follows: the process occupied by the Byzantine agent sends $0$ to process $p2$ and $1$ to process $p3$. 
Let us consider only the processes $p_2$ and $p_3$. The multiset held by $p_2$ is \{0,0,1\}. This multiset is identical with the one $p_2$ gathered in $E1$, hence its choice in $E3$ should be $0$ (identical to the one in E1). The multiset gathered by $p_3$ in $E3$ is \{1,0,1\} and identical with the one $p_3$ gathered in $E2$. Thus, $p_3$ should choose $1$ in $E3$. Execution $E3$ violates the Agreement property of Simple Approximate Agreement. This contradicts the assumption that  $\mathcal{A}$ verifies the Simple Approximate Agreement properties.
\renewcommand{\toto}{t:garayBound}
\end{proofT}


\begin{theorem}[Lower bound for Bonnet\rq{}s model]
\label{t:bonnetBound}
There is no algorithm that solves Simple Approximate Agreement in the Bonnet's model (M2) under the Mobile Byzantine faults model  if $n \leq 5f$.
\end{theorem}

\begin{proofT}
The proof follows the same general idea as the proof of Theorem \label{t:garayBound}.
Suppose that  exists an algorithm $\mathcal{A}$ verifying Simple Approximate Agreement properties in Mobile Byzantine model (M2) with $n \leq 5f$.
In all of them we consider five processes $p_0,p_1,p_2,p_3$ and $p_4$, where $p_0$ is occupied by a Byzantine agent while $p_1$ is cured (its state may be corrupted) and $p_2,p_3$ and $p_4$ are correct processes. 

Consider three executions: $E1$, $E2$ and $E3$. 
Execution $E1$ starts in a configuration where $p_2,p_3$ and $p_4$ propose $0$ while $p_1$ proposes $1$. Assume $p_0$ sends $1$ to all processes.
Each non faulty process gathers in $E1$ the multi-set \{1,1,0,0,0\} and following the Agreement and Validity properties of $\mathcal{A}$ , they have all to choose $0$ in $E1$. 

Execution $E2$ starts in a configuration where $p_2,p_3$ and $p_4$ propose $1$ while $p_1$ proposes $0$. Assume $p_0$ sends $0$ to all processes.
Each non faulty process gathers in $E2$ the multi-set \{0,0,1,1,1\} and following  the Agreement and Validity properties of $\mathcal{A}$ , they have all to choose $1$ in $E2$.

Execution $E3$ brings the contradiction. Assume that in $E3$ $p_0$ sends $0$ to $p_2$ and $1$ to $p_3$. $p2$ gathers the multiset \{1,1,0,0,0\} hence it has the same multi-set as in $E1$. $p2$ then chooses $0$. $p3$ gathers the multi-set  \{0,0,1,1,1\} and since this multi-set is identical with the one gathered in $E2$, $p3$ has to make the same choice, namely $1$. Execution $E3$ violates the Agreement property, hence  $\mathcal{A}$ do not implement the Simple Approximate Agreement.
\renewcommand{\toto}{t:bonnetBound}
\end{proofT}


\begin{theorem}[Lower bound for Sasaki\rq{}s model]
\label{t:sasakiBound}
There is no algorithm that solves Simple Approximate Agreement in the Sasaki's model (M3) under the Mobile Byzantine faults model  if $n \leq 6f$.
\end{theorem}

\begin{proofT}
The proof follows directly from the lower bound for the Simple Approximate Agreement \cite{FLM86} and the mapping defined in Section \ref{section:systemmodel}. Note that in the Sasaski\rq{}s model the number of processes with asymmetric behavior is $2f$ where $f$ is the number of Byzantine agents.
\renewcommand{\toto}{t:sasakiBound}
\end{proofT}

\begin{theorem}[Lower bound for Burhman\rq{}s model]
\label{t:burhmanBound}
There is no algorithm that solves Simple Approximate Agreement in the Burhman's model (M4) under the Mobile Byzantine faults model  if $n \leq 3f$.
\end{theorem}

\begin{proofT}
The proof follows directly from the lower bound for Simple Approximate Agreement \cite{FLM86} and the mapping defined in Section \ref{sec:background}. Note that in the Burhman's model in each round there are exactly $f$ asymmetric faulty processes.
\renewcommand{\toto}{t:burhmanBound}
\end{proofT}

%% file: conclusions.tex
This paper proves \emph{lower and upper bounds}   for achieving Approximate Agreement in the Mobile Byzantine faults model. 
Our core technique is the \emph{first mapping}  between variants of Mobile Byzantine faults models, and the Mixed-mode faults model \cite{KA94}. Our mapping then permitted to prove that the class of MSR (Mean-Subsequence-Reduce) Approximate Agreement 
algorithms are correct in the Mobile Byzantine faults model. We believe that our technique can be reused for other classical problems in Byzantine fault tolerance  (\emph{e.g.} agreement, clock synchronization, interactive consistency etc). 